\documentstyle[aps,prl,twocolumn,epsf]{revtex}
\begin{document}
\draft
\twocolumn[\hsize\textwidth\columnwidth\hsize\csname
 @twocolumnfalse\endcsname
\title{Measurements of the complex conductivity of Nb${}_{x}$Si${}_{ {1-x} }$
    alloys on the insulating side of the metal-insulator transition}
\author{Erik Helgren and George Gr\"uner}
\address{Department of Physics and Astronomy, University of California, Los Angeles}
\author{Martin R. Ciofalo, David V. Baxter, and John P. Carini}
\address{Department of Physics, Indiana University, Bloomington}
\date{\today}
\maketitle
\begin{abstract}
We have conducted temperature and frequency dependent transport measurements in amorphous Nb$_x$Si$_{1-x}$ samples in the insulating regime.  We find a temperature dependent dc conductivity consistent with variable range hopping in a Coulomb glass.  The frequency dependent response in the millimeter-wave frequency range can be described by the expression $\sigma(\omega) \propto (-\imath \omega)^\alpha$ with the exponent somewhat smaller than one.  Our ac results are not consistent with extant theories for the hopping transport.
\end{abstract}
\pacs{72.20.Ee,71.45.Gm,71.30.+h}
 \vskip2pc]
The study of conducting processes in disordered insulators has played an important role in the history of the physics of disordered materials yet retains considerable intrinsic interest.  These materials are sometimes called ``Coulomb glasses'' for the presumed complexity of the interactions between localized electrons.  Efforts to understand the conducting mechanisms in the insulating phase continue as a variety of mechanisms are possible---including thermal activation to a mobility edge, thermal activation to a neighboring localized state, and variable range hopping.  For materials well within the insulating phase, variable range hopping is likely to dominate the transport; when a delocalization transition is approached, however, it is not obvious which process will dominate \cite{imry:1995}.  In addition, the role of strong electron-electron interactions, which is known to be important in the vicinity of the metal-insulator transition \cite{hllee:1998,hllee:2000,marn:2000,massey:2000}, has not been fully explored in the insulating state.

Theories based on the variable range hopping mechanism either without strong effects due to electron-electron interactions (Mott hopping \cite{mott:1979}) or with them (Efros-Shklovskii (E-S) hopping \cite{efros:1975,efros:1985}) are well established in explaining various aspects of the dc  and low frequency dielectric response of lightly doped semiconductors and very disordered alloys.  They each produce characteristic temperature dependences:
\begin{eqnarray}
{\rm Mott:}~\sigma_{dc} &\propto& \exp[-(T_0/T)^{1/(d+1)}]\\
{\rm E-S:}~\sigma_{dc} &\propto& \exp[-(T_0/T)^{1/2}].
\label{eq:dcsigma}
\end{eqnarray}
Here $d$ is the effective spatial dimension for the hopping electrons and $T_0$ is a characteristic temperature scale that is related to the level spacing in a volume $\xi^d$ (where $\xi$ is the localization length) in the Mott theory and to the Coulomb interaction energy scale in the Efros-Shklovskii theory: $k_B T_0= e^2/\kappa \xi$ (where $\kappa$ is the dielectric constant).  Note that these expressions are valid for $T \ll T_0$.

Frequency dependent conductivity experiments are particularly useful in distinguishing between different conducting mechanisms since they directly probe the quantum state of the electrons (especially in the high frequency limit where $\hbar \omega > k_B T$).  The frequency-dependent conductivity has been measured in the disordered insulating state from audio to microwave frequencies \cite{hess:1982,migluolo:1988}, far-infrared frequencies\cite{capizzi:1980,thomas:1981}, and beyond \cite{gaymann:1993}.  For investigating the quantum behavior of the hopping conductivity, however, measurements at frequencies lower than far-infrared (below the frequency $k_B T_0/\hbar$ or the range where other excitations such as phonons are important) but still high enough to be in the quantum limit $\hbar \omega > k_B T$, {\it i.e.,} low temperature microwave and millimeter-wave conductivity experiments, are most relevant.  Here we report results for the complex conductivity in this frequency range at low temperatures for an amorphous metal-nonmetal alloy system, a-Nb$_x$Si$_{1-x}$, which exhibits a disorder-induced metal-insulator transition \cite{qhe:2000}.

Amorphous niobium-silicon alloy samples are deposited on sapphire substrates by cosputtering from separate Nb and Si sources onto rotating substrates to produce large (19 mm diameter), thick (1~$\mu$m) homogeneous samples suitable for millimeter wave transmission experiments.  The sapphire substrates have their c-axis oriented perpendicular to the plane, and are polished so that the faces are parallel.  At the same time other samples were deposited in a Hall-bar configuration for dc experiments.  Electron microprobe analysis verifies the lateral homogeneity produced by this process and was also used to estimate the niobium concentration (values are listed in Table I).

The dc conductivity was measured for each sample in a standard pumped He-4 cryostat over the temperature range 1.4--300 K.  The frequency dependent conductivity is determined from millimeter-wave transmission experiments in the frequency range 100--900 GHz and temperature range 2.8--300 K.  As previously described \cite{hllee:1998}, the transmission through the sample on the substrate oscillates as a function of frequency because of standing waves in the substrate, which acts as a Fabry-Perot resonator \cite{schwartz:1995}.  We first measure the peak frequencies for each substrate alone as a function of temperature, then remeasure after sample deposition.  The complex conductivity $\sigma_1 + \imath \sigma_2$ can be determined from the changes in peak heights and shifts in the peak frequencies \cite{born:1964}.

Figure 1 shows the dc conductivity for five Nb$_x$Si$_{1-x}$ alloy samples plotted on a logarithmic scale versus $T^{-1/2}$.  For temperatures below 10 K, the data falls on a straight line, in agreement with Eq.\ \ref{eq:dcsigma}.   The slopes in the low temperature limit give the $T_0$ values, which are summarized in Table I.  Note that in the Efros-Shklovskii theory, the value of $T_0$ becomes larger as the localization length decreases; in Ref.\ \cite{hllee:1998}, the delocalization transition occurred for Nb$_x$Si$_{1-x}$ samples with $\sigma(77~{\rm K}) > 5800~(\Omega{\rm m})^{-1}$.   Our data are not consistent with the Mott variable range hopping theory unless the effective dimension for the hopping electrons is $d=1$ (which could occur if the hopping occurred preferentially along percolating paths).

Figures 2 and 3 show the real and imaginary parts of the complex conductivity  respectively, for three samples as a function of frequency at 2.8 K.  For all three samples, the imaginary part of the conductivity is negative (capacitive) with a magnitude that is considerably larger than the real part.  Also for all three, both real and imaginary parts follow a power law frequency dependence with a similar exponent $\alpha$ for both parts.  This behavior implies that for each sample the complex conductivity can be described by:
\begin{equation}
\sigma(\omega) = \sigma_1(\omega) + \imath \sigma_2(\omega) = A \left(-\imath \frac{\omega}{\omega_o}\right)^\alpha,
\end{equation}
The fact that both the real and imaginary parts individually follow the same power law over a broad frequency range is compatible with the Kramers-Kronig relations\cite{fitnote}.  We have three techniques for determining $\alpha$ from the data: (1) fitting $\sigma_1(\omega)$, (2) fitting $\sigma_2(\omega)$, and (3) fitting $\sigma_2$ versus $\sigma_1$ (with frequency as an implicit variable) to obtain the phase angle of the complex conductivity:
\begin{equation}
\alpha = \frac{2}{\pi}\tan^{-1}\left(\frac{|\sigma_2|}{\sigma_1}\right).
\end{equation}
The three techniques give values for $\alpha$ that agree within experimental error, as shown in Table I.  The value of $\alpha$ is less than unity for all three (that is, the frequency dependence is sublinear), and the value of $\alpha$ increases as the samples become more insulating.

The relatively large values for the imaginary part of the conductivity for all three samples also implies relatively large values of the real part of the dielectric function
\begin{equation}
\epsilon_1 = -\frac{\sigma_2}{\omega \epsilon_o}
\end{equation}
as shown in the inset to Figure 3.  For all three samples, the dielectric function increases slowly as frequency decreases, and its zero frequency limit is not attained in our frequency range for Samples 1 and 2.  The samples with higher conductivities exhibit the larger values of $\epsilon_1$, which is consistent with the behavior in an insulating phase as a delocalization transition is approached---electrons in states with longer localization lengths become highly polarizable---as has been previously observed at radio frequencies in doped crystalline semiconductors \cite{hess:1982,terry:1992}.

We conclude that the electrons in these samples are in highly polarizable, strongly interacting, yet localized states.  The frequency dependence of the complex conductivity possesses a nontrivial power law frequency dependence (at least in a limited frequency range that reaches the quantum limit).  The dc conductivity follows a temperature dependence consistent with variable range hopping models, so we now turn to the predictions of those models for the high frequency conductivity.  The models can be considered to be ``two site'' models in that hopping occurs between two sites either without considering interactions at all (the Mott model) or without considering screening effects produced by virtual hops of other electrons (the E-S model).

The two variable range hopping models produce different predictions for $\sigma_1$ in the quantum regime (but not too high in frequency): $k_B T < \hbar \omega < k_B T_0$ \cite{mott:1979,efros:1985}:
\begin{eqnarray}
{\rm Mott:}~\sigma_1(\omega) &=& \pi^2 e^2 g_o^2 \hbar \omega^2 \xi^5 \ln^4(2 I_o/(\hbar \omega))\\
{\rm E-S:}~\sigma_1(\omega) &\approx&  \frac{2 \pi \epsilon_o \omega \kappa}{5 \ln(2 I_o/(\hbar \omega))}.
\label{eq:esfreq}
\end{eqnarray}
Here $g_o$ is the impurity band density of levels and $I_o$ is a microscopic energy overlap.  The E-S expression assumes a broad Coulomb gap (which would be the case for a metal-nonmetal alloy, since the level density is quite high).  Besides the different prediction for the frequency dependence (quadratic for Mott versus slightly superlinear for Efros-Shklovskii), the difference in form between the two formulas is striking and results from the universal shape of the density of levels in the Coulomb gap (which determines the absorbing transitions) in the Efros and Shklovskii theory.  The Mott expression also has a much stronger dependence on the localization length.  Note that the Efros-Shklovskii expression would be expected to depend rather weakly on the niobium concentration (through the logarithmic term containing the overlap integral value) especially since the value of $\kappa$ is assumed to be determined by the host material (amorphous silicon here), which does not include the contribution of the hopping electrons.

For $\sigma_2$, Efros \cite{efros:1985b} has produced a similar expression to Eq.\ \ref{eq:esfreq} but without the logarithmic factor, so that $\sigma_2$ should have a purely linear frequency dependence.  This would give a frequency independent value for $\epsilon_1$.

Our data for $\sigma_1$ shows neither the quadratic frequency dependence nor the strong dependence of the magnitude of $\sigma_1$ on the localization length predicted by the Mott theory ($\xi$ is inversely proportional to $T_0$ in the Mott theory for $d=1$), which rules out attempts to explain the dc  conductivity data as one-dimensional hopping of weakly interacting electrons.

The data are not fully described by the Efros-Shklovskii model.  We observe a sublinear frequency dependence for the complex conductivity (as opposed to the superlinear and linear dependences predicted for $\sigma_1$ and $\sigma_2$, respectively).  Also, the measured magnitudes of $\sigma_1$ are significantly greater than the prediction of Eq.\ \ref{eq:esfreq} if $\kappa$ is taken to be the dielectric constant of amorphous silicon (the dashed line in Figure 2 indicates the magnitude of the numerator for $\kappa=12$; the logarithmic factor in the denominator is difficult to estimate but is greater than one in the theory).  There are two caveats to consider in these comparisons, however: (1) the data may not be fully in the $T=0$ limit (since $\hbar \omega$ is only somewhat greater than $k_B T$ for our lowest frequencies and the theory is for $T=0$) and (2) for Sample 1, the value of $T_0$ is low enough that the experiment does not remain below the upper frequency limit on the range of validity for Eq.\ \ref{eq:esfreq} over our entire frequency range.

It may be interesting to consider the effects on the electrodynamics of a hopping electron in a model where screening effects of other hopping electrons are taken into account.  The host material dielectric constant $\kappa$ used in the E-S model is much smaller than the contribution to $\epsilon_1$ from the hopping electrons themselves in our frequency range for all of our samples.  Thus it is conceivable that the full dielectric response must be taken into account when screening in a Coulomb glass is considered.  Recent tunneling experiments in doped semiconductors have been interpreted as supporting the importance of many-electron composite excitations \cite{massey:2000}.

To conclude, we observe a power law dependence for the complex conductivity in the low temperature limit, implying that the hopping conductivity in our samples exhibits a richer range of behavior than can be described by the existing theories of variable range hopping.  The development of theories that account for the full dielectric response of the system and additional experiments at lower frequencies and temperatures would appear to be required to fully understand the thermal and quantum processes that control the conductivity of the Coulomb glass.

Acknowlegements---We wish to thank B. Shklovskii for useful discussions.  Research at UCLA was supported by the National Science Foundation Grant DMR-9801816.


%
%
\begin{figure}
\epsfxsize=8.5cm\epsffile{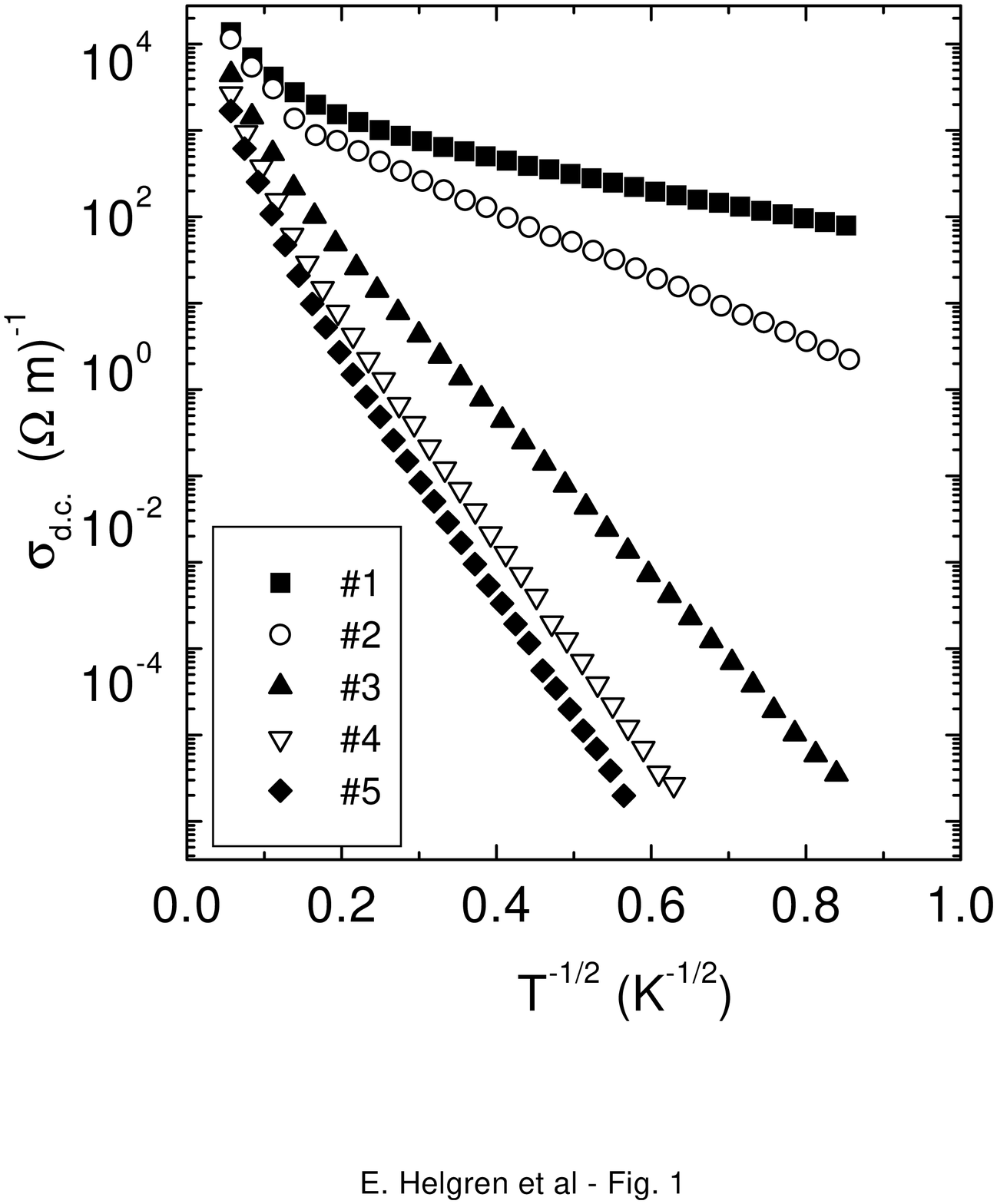}
\caption{DC conductivity
on a logarithmic scale versus $T^{-1/2}$ for a series of
Nb$_x$Si$_{1-x}$ samples.  Niobium concentrations for the samples
are given in Table I.  In the low temperature limit (below 10 K)
the data follow the form of Eq.\ 2 (the Efros-Shklovskii variable
range hopping model).} \label{fig1}
\end{figure}

\begin{figure}
\epsfxsize=8.5cm\epsffile{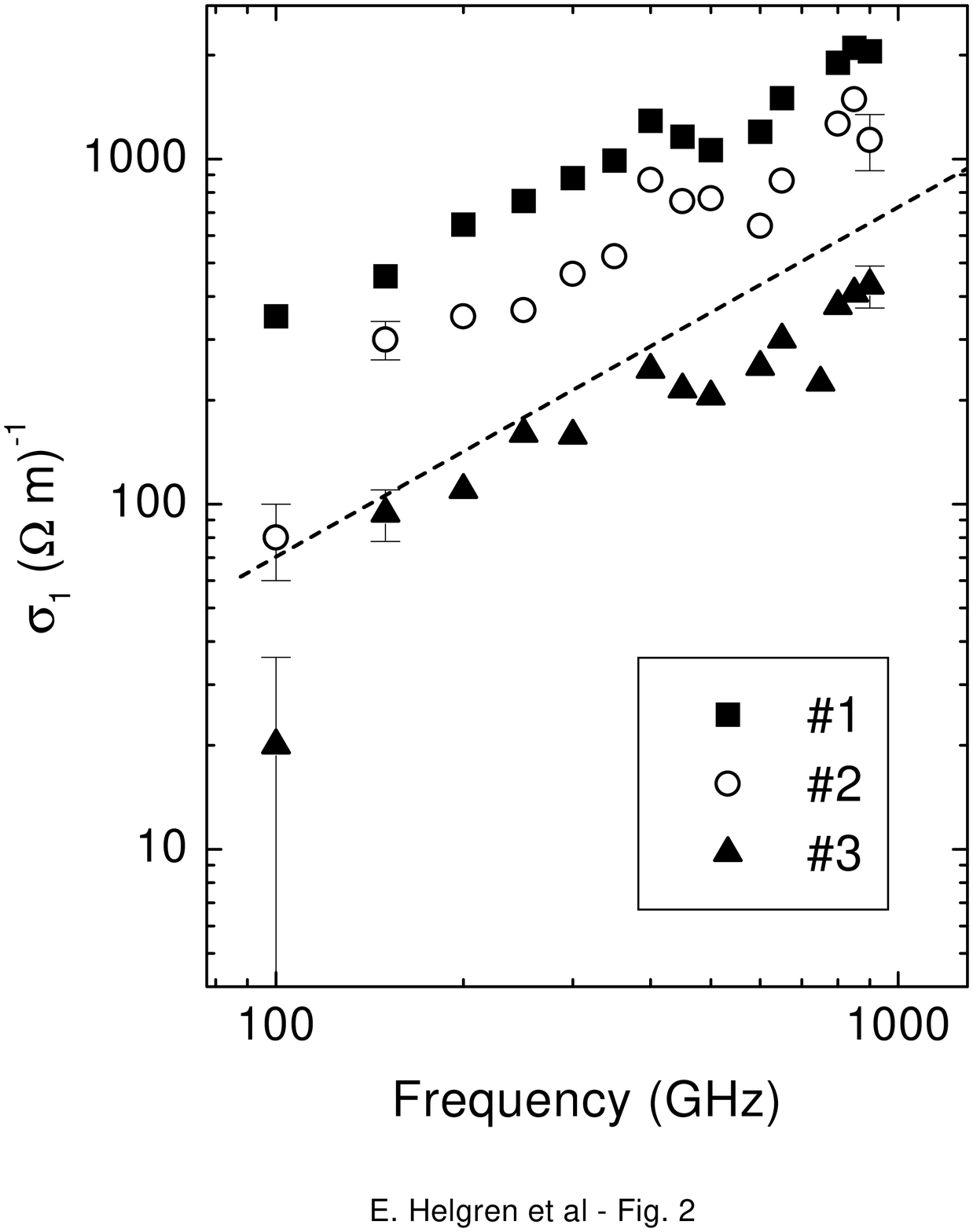} \caption{The real part of
the conductivity versus frequency (on logarithmic scales) at a
temperature of 3 K for three Nb$_x$Si$_{1-x}$ samples.  For all
three, the data can be fit by a power law frequency dependence
with an exponent somewhat less than one (the dashed line has a
slope of one).  Note that all of the measurement frequencies
exceed the frequency corresponding to the thermal energy: $k_B
T/h = 60~{\rm GHz}$.} \label{fig2}
\end{figure}

\begin{figure}
\epsfxsize=8.5cm\epsffile{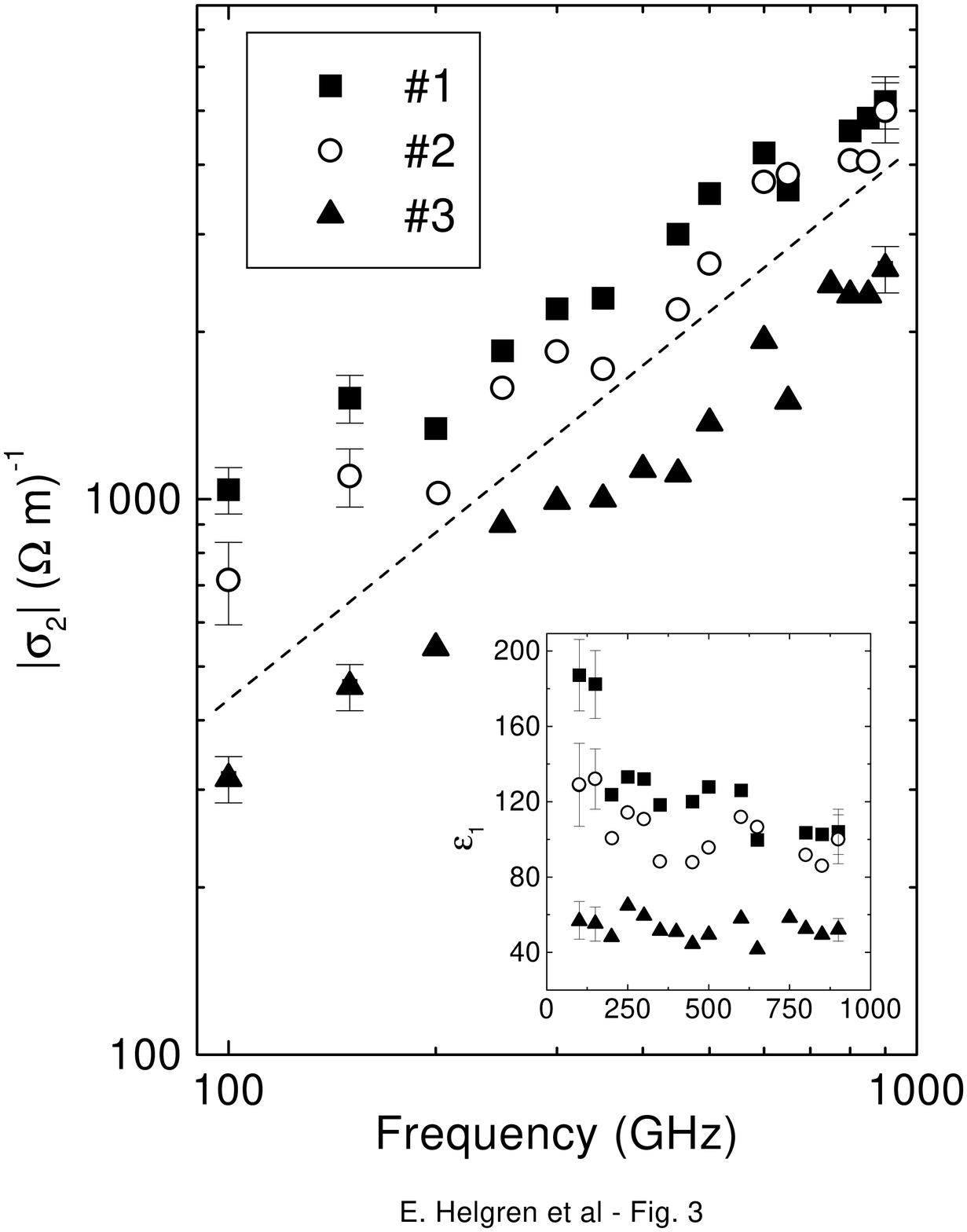} \caption{The magnitude of
the imaginary part of the conductivity versus frequency (on
logarithmic scales) at a temperature of 3 K for three
Nb$_x$Si$_{1-x}$ samples.  For all three, the sign of the
imaginary part is negative (capacitive) and its magnitude is
larger than that of the real part; the data can be fit by a power
law frequency dependence with a power that agrees with that of
the real part within experimental uncertainty.  The inset shows
the same data but expressed as the real part of the dielectric
function, which tends to increase as the frequency decreases for
all three samples.} \label{fig3}
\end{figure}
%
%
\begin{table}
\caption{NbSi alloy sample parameters.  The three $\alpha$ values result from fitting $\sigma_1$ and $\sigma_2$ to the real and imaginary parts of Eq.\ 3, respectively, and the phase angle of the complex conductivity in Eq.\ 4.  The value $A$ results from fitting the magnitude of the complex conductivity in Eq.\ 4.}
\label{samples}
\begin{tabular}{cccccccc}
Sample&$x$&$\sigma(77~{\rm K})$&$T_0$&$\alpha_{\rm fits}$&$A$\\
 &[Nb at\%]
&$[10^3 (\Omega {\rm m})^{-1}]$
&[{\rm K}]
&$\alpha_{\sigma_1},\alpha_{\sigma_2},\alpha_{{\sigma_2}/{\sigma_1}}$
&$[(\Omega{\rm m})^{-1}]$\\
 &$\pm0.5$&$\pm5\%$&$\pm2\%$&$\pm10\%$&$\pm25\%$\\
1&8.2&4.1&15.6&$0.79,0.74,0.76$&$32$\\
2&7.8&2.9&75&$0.87,0.87,0.83$&$16$\\
3&6.0&0.49&470&$0.85,0.93,0.91$&$4.6$\\
4&4.9&0.176&860&--\\
5&4.3&0.086&1010&--
\end{tabular}
\end{table}


\begin{references}
\bibitem{fitnote} Note that in a fit, the constants $A$ and $\omega_o$ are neither independent nor dimensionless:  we chose one to have convenient units ($\frac{\omega_o}{2 \pi}=1~{\rm GHz}$) and fit the other (see the Table).
\end{references}
\end{document}